\documentclass[a4paper,12pt]{article}
\usepackage[utf8]{inputenc}
\usepackage{graphicx}
\usepackage[varg]{txfonts} 
\usepackage[english]{babel}
\usepackage[margin=0.7in]{geometry}

\graphicspath{{images/}}

\begin{document}

\title{Renormalization Group in the Problem of Active Scalar Advection}
\author{N. V. Antonov$^1$ and 
M. M. Kostenko$^{1,2}$}
\date{$^1$ Department of Physics, Saint Petersburg State University,
7/9 Universitetskaya Naberezhnaya, Saint Petersburg, 199034, Russia \\
$^2$ L.~D.~Landau Institute for Theoretical Physics, Russian Academy of Sciences, 142432, Chernogolovka, Moscow Region, Russia\\
n.antonov@spbu.ru, m.m.kostenko@mail.ru}

\maketitle

\abstract{The field theoretic renormalization group (RG) is applied to the model of a near-equilibrium fluid coupled to a scalar field (like temperature or density of an impurity) which is active, that is, influencing the dynamics of the fluid itself. It is shown that the only possible nontrivial infrared (IR)
asymptotic regimes are governed by  ``passive'' fixed points
of the RG equations, where the back reaction is irrelevant. 
This result reminds of that obtained 
in [Nandy and Bhattacharjee, J. Phys. A: Math. Gen. {\bf 31}, 2621 (1998)] in a model describing active convection by fully developed turbulence. 
Furthermore, we establish the existence of ``exotic'' fixed points with negative and complex effective couplings and transport coefficients that may suggest possible directions for future studies. }

\section{Introduction} \label{sec:Intro}

Over forty years ago, Forster, Nelson and Stephen published their seminal paper in which the renormalization group (RG) method was applied to statistical theory of fluid turbulence \cite{FNS}; see also the preliminary short publication \cite{FNS1}. They studied the stochastic Navier—Stokes (NS) equation subjected to various kinds of external stirring force, stochastic Burgers equation,\footnote{Later it was reintroduced in a scalar form by Kardar, Parisi and Zhang \cite{KPZ}
and became a paradigmatic model in non-equilibrium critical phenomena.}
passively advected scalar field, and mentioned briefly an active scalar field. The paper was mostly devoted to the problem of “long tails” that arises in the derivation of hydrodynamic equation in two dimensions, but it gave a decisive impact to the RG approach in fully developed turbulence. Pioneered in \cite{GS}--\cite{AVP}, the RG theory of turbulence still remains a perspective and developing area of theoretical physics \cite{FRGNS1}--\cite{FRGNS6}. The most powerful field theoretic RG approach is reviewed in the monographs \cite{Red,Book3} and the papers \cite{UFN} -- \cite{HHL}; see also references therein.

Admittedly, so far constructive analytical approaches based on the underlying NS dynamics have had limited success in describing real turbulent flows. Most of such approaches
(like DIA or EDQMN approximation) can be viewed as one-loop approximations to certain self-consistency equations that involve infinite series of perturbative terms \cite{Orszag}.\footnote{In this respect, they are similar to the Hartree-Fock and Gor'kov equations in many-body problems, which can be interpreted as  the leading-order  approximations of certain general schemes; see, e.g., the the monograph \cite{Book}.}
The first successful attempt to work with the whole perturbative series was undertaken in the seminal paper by Wyld \cite{Wyld} and was continued, e.g., in a series of works by L'vov et. al. \cite{Lvov,Lvov2}.

An alternative approach to work with the whole perturbative series is provided by the field theoretic RG. Its important advantage is that it is based on a regular expansion in a formal small parameter and, as such, it respects the Galilean symmetry of the problem in any finite-order approximation. Furthermore, it can be naturally combined with exact functional relations (like the Ward identities and the Dyson or Schwinger equations) and the short-distance operator-product expansion; see \cite{Red}--\cite{HHL} for the reviews and the references.

One of the marker problems is the establishing of the existence of the so-called anomalous (multi-) scaling on the base of a dynamical model and calculation of the corresponding anomalous exponents within a certain perturbation expansion \cite{Frisch}.
Despite much effort, this problem remains essentially open for real turbulence, or, to be more specific, for the stochastic NS equation \cite{Lvov2}.

However, this problem was solved analytically in an exhaustive way for the celebrated Kraichnan’s rapid-change model: a scalar field advected passively by the velocity field with given statistics \cite{Kraich}; see also, e.g., \cite{FGV} for a review and references. This problem can be  accessed by a few approaches: the so-called zero-mode approach \cite{Zero2,Zero1}, numerical simulations \cite{simula}, and the RG \cite{AAV}.

The latter allows one to construct a regular perturbative expansion for the anomalous exponents, similar to famous $\varepsilon$-expansion of the critical exponents \cite{Book3,Amit,Zinn}, and to calculate the anomalous exponents to the order $\varepsilon^3$
for the rapid-change model \cite{cube} and to $\varepsilon^2$ for its generalization with the finite correlation time \cite{Ant3} and for advection by the stirred NS equation \cite{Ant7}.

The next step to access the real turbulence is to take into account the back influence of the scalar field on the dynamics of the fluid (``active scalar''). In connection to the phase transitions in binary fluids it was studied in \cite{Ruiz} -- \cite{Nandy}. 

In turbulence itself, the
comparison of the passive and scalar advection was performed, and a promising resemblance between their spectra was established \cite{Celani}--\cite{Ching}.
In particular, the so-called statistically conservation laws (on the microscopic level related to some preserved geometric configurations of convected particles) play an important role in the both cases \cite{CelV}--\cite{Ching}. But the situation is not absolutely clear yet \cite{Celani, Celani2}.\footnote{The relation between the RG, statistical conservation laws and operator-product expansion is discussed in \cite{KMG}.}

The RG approach was applied to the active scalar advection by a turbulent flow by Nandy and Bhattacharjee in \cite{Nandy}. In that situation, it is necessary to describe the velocity statistics by a full-scale dynamical model, like the stochastic NS equation (and not by a synthetic Gaussian ensemble, like the rapid-change model). They found out, that the only infrared (IR) attractive fixed point of the RG equations (and, therefore, the only possible type of IR asymptotic behavior) corresponds to a situation where the active term is IR irrelevant (in the sense of Wilson). This observation gives an explanation to previous numerical or phenomenological findings about the close resemblance between the passive and active cases.

Admittedly, the RG analysis is reliable and self-contained at small values of the expansion parameter, the exponent entering the stirring force correlation function in the NS equation (called $y$ or   
$2\varepsilon$ different studies). Extrapolation to real finite value is a separate task. And it needs to apply additional methods like the operator expansion at short distances; see \cite{Red} and the references. Strong IR divergences arise in the model at finite values of $y=2\varepsilon$, related to  most divergent contributions from the ``dangerous'' composite operators built solely by velocity fields and their time derivatives \cite{Red,UFN,JETP}.

In this paper, we present the results of the RG analysis of the active scalar field, advected by the velocity statistics introduced and studied in \cite{FNS,FNS1}. 

It corresponds to the fluid in thermal equilibrium, and, therefore, does not describe the fully developed turbulence. The advantage is that here, the problem of extrapolation to finite values of the RG expansion parameter does not exist, and perturbative results of the RG parameter can be trusted (no ``dangerous'' composite operators are expected to exist).  The active term in the NS equation is taken in the standard form, applied in \cite{Ruiz} -- \cite{Nandy}. It can be derived on the base of microscopic hydrodynamic theory and arises in the so-called model~H of equilibrium critical behaviour; see the reviews \cite{HH,FM} and chap.~5 of monograph \cite{Book3}. 

From a more phenomenological point of view, it is the simplest term that can be constructed of minimal number of fields and derivatives, and
cannot be removed by a redefinition of the pressure term.

Our main result is qualitatively the same as that of \cite{Nandy}: the only nontrivial  IR attractive fixed point corresponds to the passive scalar, where the active term in the NS equation is irrelevant.

The plan of the paper is the following.
In  section~\ref{sec:Model} we introduce the model: the NS equation with the ``active'' term and the diffusion-advection equation for scalar field, with the respective random stirred noises. Section~\ref{sec:FT} is devoted to the field theoretic formulation of the problem: the corresponding action functional 
and the elements of the Feynman diagrammatic techniques.
In the section~\ref{sec:UV} one can find the  analysis of the ultraviolet (UV) divergences: the canonical dimensions are derived
for all the fields and parameters, and possible counterterms are presented.
As a result, the multiplicative renormalizability of the model is established,
and the renormalized action functional is written down, with all the needed renormalization constants. 
In the following section~\ref{sec:One} the one-loop expressions for the relevant diagrams and all the renormalization constants are presented. 
In section~\ref{sec:RG} the RG equations are derived, and the RG functions 
(the $\beta$ functions and the anomalous dimensions) are given, with the explicit expressions for their one-loop ap\-proxi\-ma\-ti\-ons.
The section~\ref{sec:FP} lists all the fixed points of the RG equations along with the eigenvalues of the stability matrices, that determine their character as attractors of the RG equations. The last section~\ref{sec:Conc} is reserved for conclusions.

\section{The model} \label{sec:Model}

In present work we consider advection of an active scalar field by the incompressible fluid. To begin with, let us define the equation of the  scalar field advection. It has a form of usual diffusion-advection equation subjected to a random force:
\begin{equation}
\nabla_t \theta(x)   =\kappa _0 \partial^{2} \theta(x) +f(x),
\label{density}
\end{equation}
where $x = \{t, {\bf x}\}$, $\theta(x)$ is the scalar field, $\kappa_0$ is the molecular diffusivity coefficient, ${\bf v} = \{v_i(x)\}$ is the fluid velocity field,  
\begin{equation}
\nabla_t = \partial_t + v_k(x) \partial_k
\label{nabla}
\end{equation}
is the Lagrangian (Galilean covariant) derivative, 
$\partial^{2}=\partial_k\partial_k$ is the Laplace operator
(with implied summation over the repeated index),
and $f(x)$ is a Gaussian noise with a zero mean and given correlation function:
\begin{equation}
\langle f(t, {\bf x})f(t', {\bf x'}) \rangle = B_0 \delta(t-t')\delta({\bf x}-{\bf x'})
\label{noise}
\end{equation}
The amplitude $B_0>0$ will be set equal to one in the following: this can be achieved by a proper rescaling of the field $\theta$, noise and parameters. 

It is worth noting that there are two kinds of scalar fields: one is the density of a conserved quantity (for example, density of a pollutant), while the second one is a ``tracer'' (concentration of a pollutant, temperature or  enstrophy). Expression (\ref{force}) corresponds to the tracer field (no conservation is implied).
For the density field, the noise correlation function necessarily contains a derivative and therefore has the form
\begin{equation}
\langle f(t, {\bf x})f(t', {\bf x'}) \rangle = - B_0\, \delta(t-t')\,
\partial^2 \,\delta({\bf x}-{\bf x'}), \quad B_0>0.
\label{force2}
\end{equation}
 In what follows we will mostly consider the case of a tracer. The density field will be discussed later: it turns out, that the results for this case are derived in a more simple way.

The dynamics of the fluid is determined by the forced NS equation for a viscid incompressible fluid. This equation has the form:
\setlength\abovedisplayskip{5pt}
\begin{eqnarray}
\nabla_{t} v_{i} &=&
\nu_{0}\,\partial^{2}\, v_{i}
+ \partial_i \wp - \alpha_0 (\partial_i \theta)(\partial^2 \theta) + f_{i},
\label{NS} 
\end{eqnarray}
where $\wp(x)$ is the pressure, $f_i(x)$ is the external random stirring force, $\partial^2$ is the Laplacian and 
$\nabla_{t}$ is the Lagrangian derivative (\ref{nabla}).

The ``active'' term  $(\partial_i \theta)(\partial^2 \theta)$ in the NS equation 
contains two fields and three derivatives, and as such,
is the simplest nontrivial construction built of the minimal number 
of derivatives and scalar fields. (As already said, the tracer field enters the equations only in a form of spatial derivative.)
The simplest contribution of the form
$\partial_i\theta$ can be absorbed by a redefinition of the pressure $\wp$, and, therefore, does not affect the fluid dynamics.
From the other hand, this term can be derived in a microscopic approach,
within the framework of the model~H of equilibrium critical hydrodynamics; see \cite{Book3,HH,FM}. This derivation also shows that $\alpha_0>0$.

The statistics of the random force in (\ref{NS}) is chosen in the 
following way \cite{FNS,FNS1}:
\setlength\abovedisplayskip{5pt}
\begin{eqnarray}
\langle f_{i}(x) f_{j}(x') \rangle = D_0\, \delta(t-t')\, \int \frac{d{\bf k}}{(2\pi)^d}\,  k^2\, {\rm e}^{ {\rm i} {\bf k} ({\bf x}-{\bf x'})} P^{\perp}_{ij}({\bf k}).
\label{force}
\end{eqnarray}
Here the symbol $P^{\perp}_{ij}({\bf k}) = \delta_{ij} - {k_i k_j}/{k^2}$ denotes the transverse projector. The correlation function is interpreted as a thermal noise. The constant amplitude $D_0>0$ is positive, we will take a closer look on it below.

The factor of $k^2$ in (\ref{force})
respects the conservation of the fluid momentum and, simultaneously, 
ensures the fluctuation-dissipation relation. As a result, the equal-time correlation functions of the velocity field are described by the simple Maxwell's  distribution \cite{FNS}; for the detailed proof, see also~\cite{Red}.


\section{Field theoretic formulation} \label{sec:FT}
 
 According to the general theorem of De~Dominicis-Janssen \cite{Book3,Red,Zinn} the stochastic model (\ref{density}), (\ref{NS}) can be reformulated as the field-theoretic one with the doubled set of fields. 
The reformulated theory has the following action functional:
\begin{eqnarray}
{\cal S}(\Phi) &=& \frac{1}{2} v_{i}'D^{f}_{ik}  v_{k}' +
v_{i}' \left\{ -\nabla_{t} v_{i} +
\nu_{0} \partial^{2} v_{i} - \alpha_0 (\partial_i \theta)(\partial^2 \theta)\right\} +
\nonumber \\
&+& \frac{1}{2} \theta'\theta' + \theta'\left\{- \nabla_{t}\theta + \kappa _0 \partial^{2} \theta  \right\}.
\label{action}
\end{eqnarray}
Here $D^{f}_{ik}$ is the correlation function (\ref{force}),  $\Phi=\left\{v_{i}',\theta', v_{i}, \theta\right\}$ is the full set of fields which includes the usual velocity and scalar fields ($v_i$, $\theta$) and the additional auxiliary ``responce'' fields $v'$, $\theta'$. Here and below the summations over the repeated indices and the integrations over 
$x = \{t,{\bf x} \}$ are implied, for example:
\begin{equation}
 v_i' \,\partial^2\,  v_i =  \sum_{i=1}^d 
 \int\, dt\, d{\bf x}\,
 v_i' \partial^2 v_i .   
\end{equation}
The first term in (\ref{action}) comes from the correlation function (\ref{force})
and can be written as
\begin{equation}
 \frac{1}{2} v_{i}'D^{f}_{ik}  v_{k}' = \frac{1}{2}\, D_0\, 
(\partial_k v_{i}')  (\partial_k v_{i}') =
- \frac{1}{2}\, D_0\, v_{i}'\,  \partial^2\, v_{i}'.
\label{first}      
\end{equation}

This field theoretic model can be interpreted the following way: various correlation and response functions of the original stochastic problem can be represented as functional averages  over all the fields of the full set with weight $\exp S(\Phi)$. So they can be considered as the Green functions of the field theoretic model with the action functional (\ref{action}). This model corresponds to the standard Feynman diagrammatic technique with the propagators:
\begin{eqnarray}
\langle v_i v_j' \rangle &=& \langle v'_j v_i \rangle^{*} = \vcenter{\hbox
{\includegraphics [width=0.2\textwidth,clip]{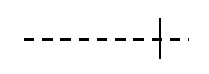}}} = \frac{P_{ij}^{\bot}({\bf k})}{- i \omega + \nu_0 k^2},
\nonumber \\
\langle v_i v_j \rangle &=& \vcenter{\hbox
{\includegraphics [width=0.2\textwidth,clip]{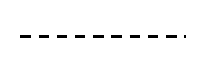}}} =  \frac{P_{ij}^{\bot}({\bf k})}{|-i\omega + \nu_0 k^2|^2},
\nonumber \\
\langle \theta \theta' \rangle &=& \vcenter{\hbox
{\includegraphics [width=0.2\textwidth,clip]{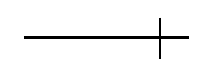}}} =  \frac{1}{{-i\omega + \kappa_0 k^2}},
\nonumber \\
\langle \theta \theta \rangle &=&  \vcenter{\hbox
{\includegraphics [width=0.2\textwidth,clip]{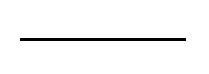}}} = \frac{1}{|-i\omega + \kappa_0 k^2|^2},
\nonumber \\
\label{lines}
\end{eqnarray}
and the three vertices:
\begin{eqnarray}
V_{ijl} &=& \vcenter{\hbox
{\includegraphics [width=0.3\textwidth,clip]{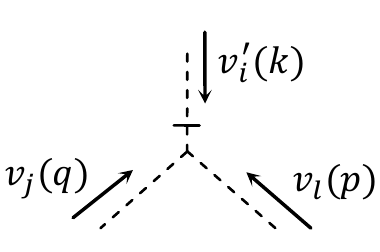}}} = i(k_l \delta_{ij} + k_j \delta_{il}),
\nonumber \\
V_{i} &=& \vcenter{\hbox
{\includegraphics [width=0.3\textwidth,clip]{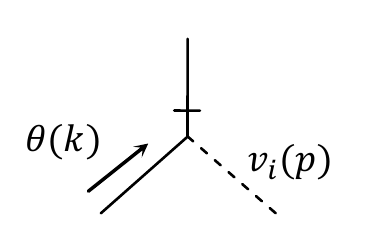}}} = - i k_i,
\nonumber \\
V_{j} &=& \vcenter{\hbox
{\includegraphics [width=0.3\textwidth,clip]{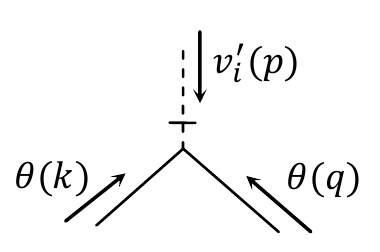}}} = i\alpha (k_i q^2 + q_i k^2)  = - i\alpha (k_i(pq) + q_i (pk)) .
\nonumber \\
\label{verticies}
\end{eqnarray}

The role of the coupling constants is played by the parameters $g_0$, $w_0$ and $u_0$ defined by the relations 
\begin{equation}
D_0 = g_0\, \nu_0^3, \quad
\alpha_0 = w_0\, \nu_0^3, \quad
\kappa_0 = u_0 \nu_0,
\label{D0}
\end{equation}.
The dimensional analysis (see the next section) shows that the first two coupling
constants scale as $g_{0}, w_0 \sim \Lambda^{\varepsilon}$, where $\Lambda$ is a characteristic UV momentum, while $u_0$ is dimensionless. Thus, the model as a whole becomes logarithmic (all the couplings become dimensionless) 
at  $\varepsilon \equiv 2-d = 0$, and the UV divergences have the form of  poles in $\varepsilon$ in the Green functions. Although $u_0$ is not an expansion parameter, it should be treated on the same footing as $g_{0}, w_0$, because the renormalization constants and the RG functions will depend on it.

 \section{UV divergences and renormalization} \label{sec:UV}
 
 It is well known that the analysis of UV divergences is based on the
analysis of canonical dimensions; see, {\it e.g.,} \cite{Book3,Amit,Zinn}.
Dynamical models like (\ref{action}) have
two independent scales: the time scale $T$ and the length scale $L$. Thus
the canonical dimension of any quantity $F$ (a field or a parameter) is
described by two numbers, the
frequency dimension $d_{F}^{\omega}$ and the momentum dimension $d_{F}^{k}$,
defined such that $[F] \sim [T]^{-d_{F}^{\omega}} [L]^{-d_{F}^{k}}$.
The obvious consequences of the definition are the relations
\begin{eqnarray}
d_k^k=-d_{\bf x}^k=1,\quad d_k^{\omega} =d_{\bf x}^{\omega }=0,
\quad
d_{\omega }^k=d_t^k=0, \quad d_{\omega }^{\omega }=-d_t^{\omega }=1.
\label{canon}
\end{eqnarray}
The other dimensions are found from the requirement that each term of the
action functional be dimensionless (with respect to the momentum and the
frequency dimensions separately). Then one introduces the total canonical
dimension \cite{Book3}
\begin{eqnarray}
d_{F}=d_{F}^{k}+2d_{F}^{\omega},
\label{total}
\end{eqnarray}
which plays in the theory of renormalization of dynamical models the same
part as the conventional canonical dimension does in static problems.
The canonical dimensions for the model (\ref{action}) are given in
table~1, including renormalized parameters (without the subscript ``o''),
which will be introduced later on.

\begin{table*}
\caption{Canonical dimensions of the fields and parameters in the
models}
\label{table1}
\begin{tabular}{c|c|c|c|c|c|c|c|c|c}
$F$ & $v'$ & $v$ & $ \theta' $ &  $\theta$ &
$m$, $\mu$, $\Lambda$ & $\nu_0$, $\nu$  & $\alpha_0$, $\alpha$ & $g_0,w_0$
& $u_0,g,w,u$
\\
\hline
$d_{F}^{k}$ & $d+1$ & $-1$ & $d/2$ & $d/2$ & $1$  & $-2$
& $-d-4$ &  $2-d$ & $0$ \\
$d_{F}^{\omega}$ & $-1$ & 1 & 1/2 & $-1/2$ & 0 & 1 &
3 & 0 & 0\\
$d_{F}$ & $d-1$ & 1 & $d/2+1$ & $d/2-1$ & $1$ & $0$ & $2-d$ & $2-d$ &0 \\
\end{tabular}
\end{table*}

From table~1 it follows that, as already mentioned, the model becomes logarithmic (the coupling constants become dimensionless) at  $\varepsilon = 2-d = 0$, and the UV divergences have the form of  poles in $\varepsilon$ in the Green functions.

The total canonical dimension of any 1-irreducible Green function $\Gamma$
(the formal index of UV divergence) is given by the expression
\begin{eqnarray}
\delta_{\Gamma} = d+2 - \sum_{\Phi} N_{\Phi} d_{\Phi},
\label{index}
\end{eqnarray}
in the logarithmic theory (that is, at $\varepsilon=0$).
Here $N_{\Phi}$ are the numbers of the fields entering into the function
$\Gamma$, $d_{\Phi}$ are their total canonical dimensions, and the summation
over all types of the fields $\Phi$ is implied. Superficial UV
divergences, whose removal requires counterterms, can be present only in
the functions $\Gamma$ with a non-negative integer $\delta_{\Gamma}$.
The counterterm is a polynomial in frequencies and momenta of degree
$\delta_{\Gamma}$, with the convention that $\omega \sim k^{2}$.

Dimensional analysis should be augmented by the following considerations.

If, for some reason, a number of external momenta occurs as an overall factor in all diagrams of a certain 1-irreducible Green function, the real index of divergence should be properly reduced.

In the present case the fields $\theta'$ and $v'$ enter all the vertices in the form of spatial derivatives, see (\ref{action}). Thus, any appearance of $\theta'$ or $v'$ in a certain 1-irreducible function gives an external momentum, and the real index of divergence takes on the form
\begin{eqnarray}
\delta'_{\Gamma} = \delta_{\Gamma} - N_{\theta'}  - N_{v'}.
\label{realindex}
\end{eqnarray}

As a manifestation of causality, all the 1-irreducible diagrams without external “tails” of the response fields $v', \theta'$ contain self-contracted circuits of retarded propagators and therefore vanish. Thus, it is sufficient to consider only functions with $N_{v'}+N_{\theta'} \ge 1$.

The action functional  (\ref{action}) is invariant with respect to the simultaneous reflection of the scalar fields, $\theta \to - \theta$, $\theta' \to - \theta'$. Therefore all the Green functions with odd total number of the scalar fields vanish (no diagrams can be constructed). In particular, this excludes the 1-irreducible function $\langle \theta' \theta \theta \rangle$ and the corresponding counterterm $\theta' (\partial \theta)^2$.

The counterterms having the form of total derivatives (or reduced to such form using the integration by parts) vanish after the integration over $x$ and should be ignored; consequently, the counterterms that differ by a total derivative should be identified.

Of course, the transversality conditions $\partial_i v_i = \partial_i v'_i = 0$ for the vector fields should not be forgotten.

The analysis shows that our model is multiplicatively renormalizable: all the counterterms required for the elimination of UV divergences can be reproduced by rescaling of the terms already present in the action (\ref{action}); what is more, some terms need no rescaling (some counterterms allowed by dimensional analysis are forbidden by additional considerations discussed above). The needed counterterms are
$\partial v'\partial v'$, $v'\partial^2 v$, $v'(\partial \theta)\partial^2\theta$, $\theta'\partial^2 \theta$. The counterterms
$v'\nabla_t v$, $\theta' \nabla_t \theta$, $\theta'\theta'$ are forbidden
by Galilean symmetry and/or by the real index.

The resulting renormalized action functional has the form:
\begin{eqnarray}
{\cal S}_R(\Phi) &=& \frac{1}{2} Z_D\,D\, \partial_k v_{i}' \partial_k v_{i}' +
v_{i}' \left\{ -\nabla_{t} v_{i} +
Z_\nu\, \nu \partial^{2} v_{i} \right\} - 
\nonumber \\
&-& v_{i}'  Z_\alpha\, \alpha\,  (\partial_i \theta)(\partial^2\theta)+
+ \frac{1}{2} \theta'\theta' + \theta'\left\{ - \nabla_{t}\theta  + Z_\kappa\, \kappa \partial^{2} \theta   \right\},
\label{Sr}
\end{eqnarray}
where the dimensionless renormalization constants $Z_i$ are naturally reproduced as multiplicative renormalization of the parameters:
\begin{eqnarray}
D_0 = D \, Z_{D}, \quad 
\nu_0 =  \nu  Z_{\nu}, 
\quad
\alpha_0 =   \alpha \, Z_{\alpha},
\quad
\kappa_0 =  \kappa  Z_{\kappa},
 \end{eqnarray}
where $D,\nu,\alpha,\kappa$ are renormalized parameters and the reference mass $\mu$ is an additional parameter of the renormalized theory. From now on, we assume that the minimal subtraction (MS) scheme of renormalization is used.

To pass to dimensionless couplings $g,w,u$
we make the substitutions, cf. eqn.~(\ref{D0}):
\begin{equation}
D = g \mu^{\varepsilon} \nu^3, \quad \alpha = w \mu^{\varepsilon}  \nu^3, \quad \kappa = u \nu,
\end{equation}
which leads to the following relations for $Z_i$:
\begin{equation} 
Z_g = Z_{D}\,  Z_{\nu}^{-3}, 
\quad
Z_w = Z_{\alpha}\,  Z_{\nu}^{-3},
\quad
Z_u = Z_{\kappa}  \, Z_\nu^{-1} .
\label{relatio}
\end{equation}
No renormalization of the fields is required.

\section{One-loop results} \label{sec:One}

In order to calculate the renormalization constants $Z_i$ in the one-loop approximation one has to calculate UV divergent parts of the diagrams shown below. We do not give details of the calculation, which is rather standard for dynamical models; see, e.g., Appendix in  \cite{Ca1} and sec.~C in \cite{Ca2}. Below we give the coefficients in front of the pole $1/\varepsilon$ in the diagrams. 

For $Z_D$ -- 1-irreducible function $\langle v'_iv'_j\rangle$: 

\begin{figure}[h]
\begin{minipage}[h]{0.49\linewidth}
\center{\includegraphics[width=1\linewidth]{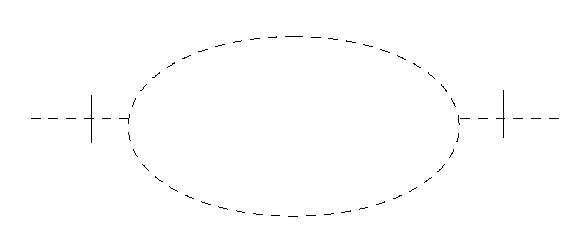}}
\end{minipage}
 $\displaystyle{ = \frac{g^2\nu^3}{8}\, p^2\, P^{\perp}_{ij}({\bf p}) * \frac{1}{2}}$
\hfill

\begin{minipage}[h]{0.49\linewidth}
\center{\includegraphics[width=1\linewidth]{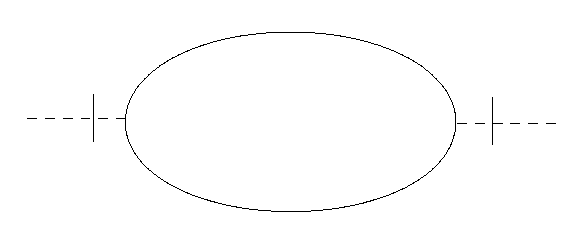}}
\end{minipage}
 $\displaystyle{ = \frac{w^2\nu^3}{8u^3}\, p^2\, P^{\perp}_{ij}({\bf p}) * \frac{1}{2}}$
\end{figure}

\newpage

For $Z_{\nu}$ -- 1-irreducible function $\langle v'_iv_j\rangle$:

\begin{figure}[h]
\begin{minipage}[h]{0.49\linewidth}
\center{\includegraphics[width=1\linewidth]{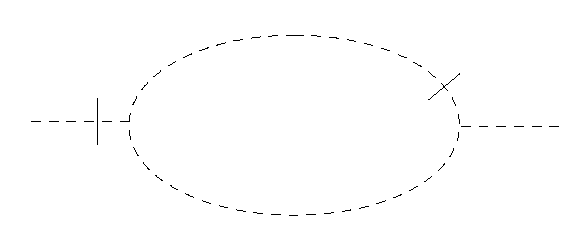}}
\end{minipage}
 $\displaystyle{ = \frac{-g\nu}{16}\,p^2\, P^{\perp}_{ij}({\bf p})}$.
\hfill

\begin{minipage}[h]{0.49\linewidth}
\center{\includegraphics[width=1\linewidth]{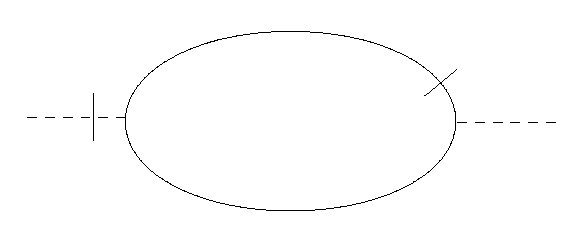}}
\end{minipage}
 $\displaystyle{ = \frac{-w\nu}{16u^2}\, p^2\, P^{\perp}_{ij}({\bf p})}$.
\end{figure}


For $Z_{\kappa}$ -- 1-irreducible function 
$\langle \theta' \theta\rangle$:

\begin{figure}[h]
\begin{minipage}[h]{0.49\linewidth}
\center{\includegraphics[width=1\linewidth]{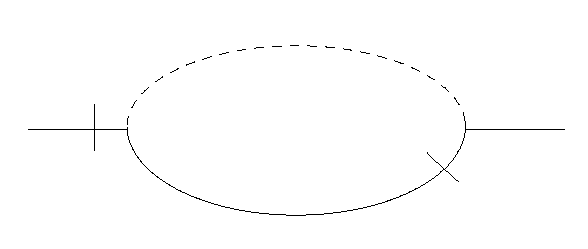}}
\end{minipage}
 $\displaystyle{ = \frac{-g\nu}{4(u+1)}\,p^2}$.
\hfill

\begin{minipage}[h]{0.49\linewidth}
\center{\includegraphics[width=1\linewidth]{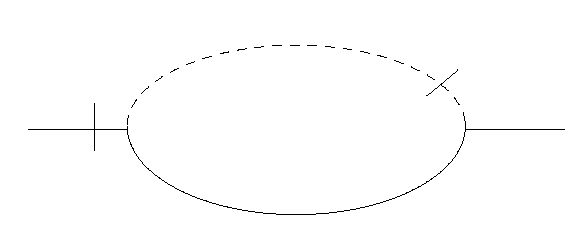}}
\end{minipage}
 $\displaystyle{ = \frac{w\nu}{4u(u+1)}\,p^2 }$.
\end{figure}

\newpage 

For $Z_{\alpha}$ -- -- 1-irreducible function 
$\langle v'_j\theta\theta\rangle$:

\begin{figure}[h]
\begin{minipage}[h]{0.2\linewidth}
\center{\includegraphics[width=1\linewidth]{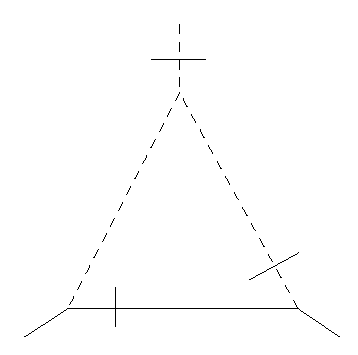}}
\end{minipage}
 $\displaystyle{ =- iw\,\nu^3\,[p_j(p+q,q)+q_j(p+q,p)]\, 
 \frac{g}{16 (u+1)}}$
\hfill

\begin{minipage}[h]{0.2\linewidth}
\center{\includegraphics[width=1\linewidth]{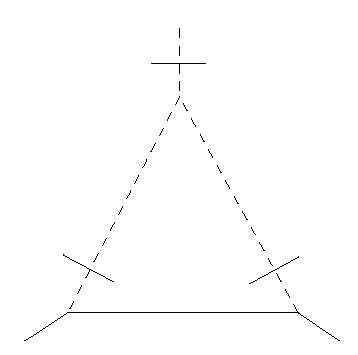}}
\end{minipage}
$\displaystyle{ = iw\,\nu^3\,[p_j(p+q,q)+q_j(p+q,p)]\, \frac{w}{16u(u+1)}}$
\hfill

\begin{minipage}[h]{0.2\linewidth}
\center{\includegraphics[width=1\linewidth]{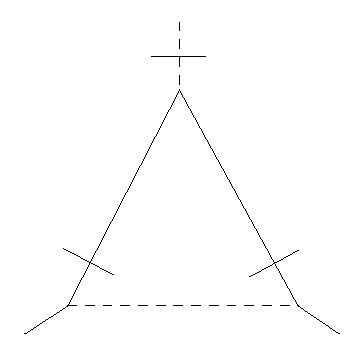}}
\end{minipage}
$\displaystyle{ = - iw\,\nu^3\,[p_j(p+q,q)+q_j(p+q,p)]\, \frac{g}{16u(u+1)}}$
\hfill

\begin{minipage}[h]{0.2\linewidth}
\center{\includegraphics[width=1\linewidth]{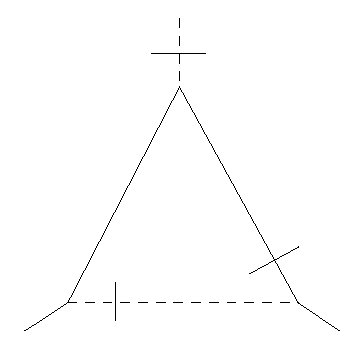}}
\end{minipage}
$\displaystyle{ = iw\,\nu^3\,[p_j(p+q,q)+q_j(p+q,p)]\, \frac{w}{16(u+1) u^2}}$
\end{figure}

\newpage
In the first two diagrams additional factor $* 1/2$ is the symmetry coefficient.
Note that the contribution of the second diagram for $Z_\kappa$ is positive, in contrast to the three preceding diagrams. This means that the contribution of the active term reduces the effective diffusivity coefficient, which is not typical for models of advection. The same effect takes place for the model studied by~\cite{Nandy}.

After calculating the one-loop diagrams one can find the expressions for $Z's$ which are necessary to eliminate the divergences: the Green functions of the renormalized model should be UV finite in the limit $\varepsilon \to 0$. Using the MS scheme, one obtains the following results:
\begin{eqnarray}
Z_D &=& 1 -  \frac{g}{32 \pi \varepsilon } -  \frac{w^2}{ 32\pi\, g\, u^3  \varepsilon }, \nonumber \\
Z_{\nu} &=& 1 - \frac{g}{32 \pi  \varepsilon } - \frac{w}{32\pi\, u^2  \varepsilon }, \nonumber \\
Z_{\kappa} &=& 1 - \frac{g}{8 \pi u(u+1)  \varepsilon } + \frac{w}{8 \pi\, u^2(u+1)  \varepsilon },
\nonumber \\
Z_{\alpha} &=& 1 + \frac{1}{32 \pi (u+1)  \varepsilon } \left\{ g + \frac{g}{u} - \frac{w}{u^2} - \frac{w}{u} \right\},
\label{Zs}
\end{eqnarray}
with the corrections of higher orders in $g$ and $w$.
The other constants are found from the relations~(\ref{relatio}).

\section{RG equations} \label{sec:RG}

Let us pass to derivation of the RG equations; detailed
discussion can be found in  \cite{Book3,Amit,Zinn}. 

Consider the renormalized correlation function $G^{R} =\langle \Phi\cdots\Phi\rangle_{R}$. Owing to the absence of the fields renormalization, it differs from the original (unrenormalized) one $G =\langle \Phi\cdots\Phi\rangle$ only by the choice of parameters. 
Indeed, the relation
${\cal S}_{R} (\Phi,e,\mu) = {\cal S} (\Phi,e_{0})$ between the
functionals (\ref{action}) and (\ref{Sr}) results in the relations
\begin{equation}
G(e_{0},\dots) = G^{R}(e,\mu,\dots)
\label{multi}
\end{equation}
between the correlation functions. Here $N_{\phi}$ and $N_{\phi'}$ are the numbers of corresponding fields
entering into $G$;
$e_{0}=\{\nu_0, g_{0}, u_{0}, v_{0} \}$  is the full set of
bare parameters and $e=\{ \nu, g, u, v  \}$ are their renormalized
counterparts; the ellipsis stands for the other arguments
(times, coordinates, momenta {\it etc.}).

We use $\widetilde{\cal D}_{\mu}$ to denote the differential operation
$\mu\partial_{\mu}$ for fixed $e_{0}$ and operate on both sides of the
equation (\ref{multi}) with it. This gives the basic RG differential
equation:
\begin{equation}
{\cal D}_{RG} \,G^{R}(e,\mu,\dots) = 0.
\label{RG1}
\end{equation}
Here  ${\cal D}_{RG}$ is the operation $\widetilde{\cal D}_{\mu}$
expressed in terms of the renormalized variables:
\begin{equation}
{\cal D}_{RG}= {\cal D}_{\mu} + \beta_{g}\partial_{g} +
\beta_{u}\partial_{u} + \beta_{w}\partial_{w}
- \gamma_{\nu}{\cal D}_{\nu} .
\label{RG2}
\end{equation}
Here we have written ${\cal D}_{s} \equiv s\partial_{s}$ for any variable
$s$. The anomalous dimension $\gamma_{e}$ of a certain parameter $e$
is defined as
\begin{equation}
\gamma_{e}= Z_{e}^{-1} \widetilde{\cal D}_{\mu} Z_{e} =
\widetilde{\cal D}_{\mu} \ln Z_e ,
\label{RGF1}
\end{equation}
and the $\beta$ functions for the three dimensionless coupling
constants $g$, $u$ and $w$ are
\begin{eqnarray}
\beta_{g} &=& \widetilde {\cal D}_{\mu} g = g\,[-\varepsilon-\gamma_{g}],
\nonumber \\
\beta_{w} &=& \widetilde {\cal D}_{\mu} w = w(- \varepsilon - \gamma_w),
\nonumber \\
\beta_{u} &=& \widetilde {\cal D}_{\mu} u =  -u\gamma_{u},
\label{betagw}
\end{eqnarray}
where the second equalities result from the definitions and the
relations (\ref{relatio}).

In the MS scheme all the renormalization constants have the form
\begin{eqnarray}
Z_{e} = 1 + \sum_{n=1}^{\infty} z^{(n)} \varepsilon^{-n},
\label{ZMS}
\end{eqnarray}
where the coefficients $z^{(n)}$ do not depend on $\varepsilon$. Then from
the definition and the expressions (\ref{betagw}) it follows that
the corresponding anomalous dimension is determined solely by the
first-order coefficient:
\begin{eqnarray}
\gamma_{e}=-{\cal D}_{g} z^{(1)},
\label{E}
\end{eqnarray}
see, {\it e.g.,} the monograph \cite{Book3}. Thus, from expressions (\ref{Zs}) we obtain the following one-loop answers for the anomalous dimensions:
\begin{eqnarray}
\gamma_D &=& \gamma_g + 3\gamma_{\nu} =    \frac{g}{32 \pi} +  \frac{w^2}{g32 \pi u^3},
\nonumber \\
\gamma_{\nu} &=&  \frac{g}{32 \pi} + \frac{w}{32 \pi u^2},
\nonumber \\
\gamma_{\kappa} &=& \gamma_u + \gamma_{\nu} = \frac{g}{8 \pi u(u+1)} - \frac{w}{8 \pi u^2(u+1)},
\nonumber \\
\gamma_{\alpha} &=& \gamma_w + 3\gamma_{\nu} = - \frac{1}{32 \pi (u+1)}
\left\{ g + \frac{g}{u} - \frac{w}{u^2} - \frac{w}{u}\right\},
\cite{gammas}
\end{eqnarray}
with the higher-order corrections in $g$ and $w$.

Then it is easy to find all the $\beta$ functions. To simplify the expressions, in the following we introduce the variables $\hat{g} = g/(2\pi)$ and $\hat{w} = w/(2\pi)$. Then the one-loop answers for $\beta$ functions are:
\begin{eqnarray}
\beta_u &=& -u\gamma_u = u(\gamma_{\nu} - \gamma_{\kappa})= -\frac{1}{4(u+1)} \left( \hat{g}-\frac{\hat{w}}{u}\right)+ \frac{u}{16}\left( \hat{g}+\frac{\hat{w}}{u^2}\right), \label{bu} \\
\beta_w &=& -\varepsilon w - w\gamma_w = -\varepsilon w + w(3\gamma_{\nu} - \gamma_{\alpha}) \nonumber \\
{} &=&  -\varepsilon w + \frac{w}{16(u+1)} \left( \hat{g} + \frac{\hat{g}}{u} - \frac{\hat{w}}{u^2} - \frac{\hat{w}}{u} \right) + \frac{3w}{16} \left( \hat{g} + \frac{\hat{w}}{u^2}\right), \label{bw} \\
\beta_g &=& -\varepsilon g - g \gamma_g = -\varepsilon g + g (3\gamma_{\nu}-\gamma_1) \nonumber \\
{} &=& -\varepsilon g - g \left(\frac{\hat{g}}{16}+ \frac{\hat{w}^2}{16\hat{g}u^3}\right) + 3g\left(\frac{\hat{g}}{16} + \frac{\hat{w}}{16u^2}\right). 
\label{bg}
\end{eqnarray}


 \section{Fixed points} \label{sec:FP}

Possible IR asymptotic regimes of a renormalizable
field theoretic model are defined by IR attractive fixed points of the
corresponding RG equations. The coordinates $g_{*}$
of the fixed points are found from the equations
\begin{equation}
\beta_{i} (g{*}) =0,
\label{points}
\end{equation}
where $g=\{g_i\}$ is the full set of coupling constants and $\beta_{i}$
is the full set of the $\beta$ functions.
The type of a fixed point is determined by the matrix
\begin{equation}
\Omega_{ij}=\partial\beta_{i}/\partial g_{j}|_{g=g_{*}}.
\label{Omega}
\end{equation}
For the IR attractive fixed points the matrix $\Omega$ is positive, that is, the real parts of their eigenvalues are positive.

In our model, $g = \{g, u, w\}$, and the $\beta$ functions are given
by the relations (\ref{bu})-(\ref{bg}). Analysis of these expressions 
reveals the following fixed points:
\mbox
1) The line of Gaussian (free) fixed points:
\begin{equation}
g_* = w_* =0, \quad u_*\ {\rm arbitrary},
 \end{equation}
with the  eigenvalues:
\begin{equation}
-\varepsilon, \quad -\varepsilon, \quad 0.
\end{equation}
This line is IR attractive for $\varepsilon<0$ ($d>2$). All the nonlinearities are unimportant in the IR region, as can be anticipated from the analysis of canonical dimensions. The vanishing of one of the eigenvalues reflects the fact that this line is indifferent to the change of the ratio $u=\kappa/\nu$: the equations for the velocity and the scalar field decouple.

2a and 2b. Passive fixed points:
\begin{equation}
w_* = 0, \quad u_* = \frac{-1\pm\sqrt{17}}{2}, \quad \hat{g}_* = 8\varepsilon; 
\label{missp}
\end{equation}
The eigenvalues:
\begin{equation}
\varepsilon, \quad \frac{9}{16}\varepsilon + \frac{\sqrt{17}}{16}\varepsilon, \quad \frac{17}{16}\varepsilon - \frac{\sqrt{17}}{16}\varepsilon,
\end{equation}
and
\begin{equation}
\varepsilon, \quad \frac{9}{16}\varepsilon - \frac{\sqrt{17}}{16}\varepsilon, \quad \frac{17}{16}\varepsilon + \frac{\sqrt{17}}{16}\varepsilon,
\end{equation}
IR attractive for $\varepsilon>0$ ($d<2$). The active term is irrelevant, and one returnes to the ordinary passive advection.\footnote{In the expression (3.69) in \cite{FNS} that corresponds to ours $u_*$ in (\ref{missp}) 
there is a misprint. It is also worth mentioning that the authors call this expression ``intriguing''.}

3. The ``active'' fixed point
\begin{equation}
u_*=-3, \quad \hat{w}_* = \frac{72\varepsilon}{5} , \quad \hat{g}_* = \frac{24\varepsilon}{5} 
\end{equation}
with the eigenvalues:
\begin{equation}
\varepsilon, \quad \frac{7}{20}\varepsilon + \frac{\sqrt{145}}{20}\varepsilon, \quad \frac{7}{20}\varepsilon - \frac{\sqrt{145}}{20}\varepsilon,
\end{equation}
which is always unstable (the eigenvalues have different signs).

4a and 4b. Complex ``active'' fixed points:
\begin{eqnarray}
 \nonumber u_* &=& i \sqrt{7}, \\ 
\hat{w}_* &=& \frac{56 (- 11 + i\sqrt{7})}{3 + 7 \sqrt{7} i}\, \varepsilon = \frac{28 \varepsilon}{11} + i\frac{140 \sqrt{7}\varepsilon}{11}, \\
\hat{g}_* &=&  \frac{8(5 + i\sqrt{7})\sqrt{7}}{3 + 7\sqrt{7}i}\, \varepsilon = \frac{56 \varepsilon}{11} + i \frac{16 \sqrt{7} \varepsilon}{11}  \nonumber
\end{eqnarray}
with the eigenvalues:
\begin{equation}
\varepsilon, ~~~ -\frac{i \varepsilon}{352}(- 11\sqrt{17} + 33i + \sqrt{-38770 - 15958i\sqrt{7}}), ~~~ -\frac{i \varepsilon}{352}(- 11\sqrt{17} + 33i - \sqrt{-38770 - 15958i\sqrt{7}}),
\end{equation}
or  approximately:
\begin{equation}
\varepsilon, ~~~ (-0.5289574958 - 0.1909278136 i)\varepsilon, ~~~ (0.7164574958 + 0.3562872705 i)\varepsilon.
\end{equation}
And, finally, the complex conjugate point
\begin{eqnarray}
 \nonumber u_* &=& - i \sqrt{7}, \\ 
\hat{w}_* &=& \frac{56 (11 + i\sqrt{7})}{-3 + 7 \sqrt{7} i}\, \varepsilon = \frac{28 \varepsilon}{11} - i\frac{140 \sqrt{7}\varepsilon}{11}, \\
\hat{g}_* &=&  \frac{8(5 - i\sqrt{7})\sqrt{7}}{-3 + 7\sqrt{7}i}\, \varepsilon = \frac{56 \varepsilon}{11} - i \frac{16 \sqrt{7} \varepsilon}{11}  \nonumber
\end{eqnarray}
with the eigenvalues:
\begin{equation}
\varepsilon, \quad -\frac{i \varepsilon}{352}(11\sqrt{17} + 33i + \sqrt{-38770 + 15958i\sqrt{7}}), \quad -\frac{i \varepsilon}{352}( 11\sqrt{17} + 33i - \sqrt{-38770 + 15958i\sqrt{7}}),
\end{equation}
or, approximately:
\begin{equation}
\varepsilon, \quad (-0.5289574958 + 0.1909278136 i)\varepsilon, \quad (0.7164574958 - 0.3562872705 i)\varepsilon.
\end{equation}
These two points are always unstable (the real parts of the eigenvalues have different signs for all $\varepsilon$).

For $\varepsilon=0$ ($d=2$) there is a line of fixed points $g_*=w_*=0$, $u_*$ arbitrary. Direct numerical integration of the differential equations for the running coupling constants shows that this point is IR attractive for the physical initial data $g>0$, $w>0$, at least for some interval of values for $u$. The sample RG flow for the special case $u=1$ is depicted in Figure~1. 

\begin{figure}[h!]
\center
\includegraphics [width=0.8\textwidth,clip]{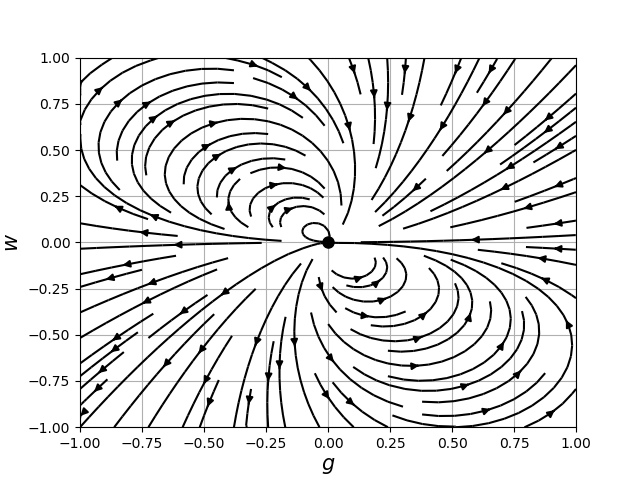}
\caption{The RG flow in the $g$--$w$ plane for $u=1$ and $d=2$}
\label{fig:Plain}
\end{figure}

Let us conclude with a brief discussion of the density model with the correlation function (\ref{force2}). Again, we can set $B_0=1$. Then the analysis of the canonical dimensions gives
\[ d^k_\theta = d/2+1, \quad d^k_{\theta'}= d/2-1, \quad 
d^k_{\alpha_0} = -d-6, \quad d^k_{w_0} = -d \]
for the momentum dimensions and
\[ d_\theta =  d_{\theta'}= d/2, \quad 
d_{\alpha_0} = d_{w_0} = -d \]
for the total dimensions. All the other dimensions remain the same as in table~1.

From these results it follows that the total dimension of the coupling constant $w_0$ is negative for all $d>0$, so that the active term appears IR irrelevant already at the level of simple dimensional analysis: no refined RG analysis is needed\footnote{The relevant dimensionless parameter $w_0 k^d$ vanishes in the IR asymptotic region $k\to0$.}; 
the fact also mentioned by~\cite{FNS}.

\section{Conclusion} \label{sec:Conc}

At first sight, our results look rather disappointing. The IR attractive fixed point is trivial (Gaussian) for $d>2$ and passive for $d<2$. The back reaction of the scalar field on the velocity dynamics is always negligible in the leading term of the IR asymptotic scaling behaviour. But this result appears robust in the sense that it agrees with the result derived earlier by Nandy and Bhattacharjee who considered another type of the random noise, corresponding to advection by a turbulent fluid \cite{Nandy}.

In agreement with \cite{Nandy}, the nonlinearity gives contributions of different signs into effective viscosity and diffusion coefficients. This means that the back reaction of the scalar field would produce the mechanical energy from the thermodynamic one. Probably this is the physics reason why the active IR regime is not realized as an attractive RG fixed point.

To avoid possible misunderstanding, it should be stressed that the ``active'' term remains to be present in the model: it can strongly affect non-universal quantities (e.g., amplitudes in scaling laws, like Kolmogorov or Batchelor constants) and correction terms to the leading IR behaviour (which can quantitatively be essential). Probably, these results give an explanation to the similarity between the spectra of passive and active fields, in spite of the serious  differences between the underlying motion of the impurity particles \cite{Celani}-\cite{Ching}.

Furthermore, the apparently ``unphysical'' fixed points 3 and 4 may have a sound physical interpretation. 

The negative value $u_*<0$ means that one of the effective viscosity or, more likely, diffusivity coefficients tend to negative values at some intermediate scales. 
A similar effect was discussed a long ago in a number of studies of non-equilibrium stochastic models \cite{Yakhot}. Along with the instability of the corresponding RG fixed points, this suggests that the original model itself might be incomplete and may be modified by adding the higher-order terms. 

The complex value of the viscosity coefficient (or better to say, of its effective analog) was encountered for stochastic models of Langmuir plasma turbulence \cite{plasma,plasma1} and for the stochastic version of the nonlinear Schr\"{o}dinger equation \cite{Tauber}. 

These considerations suggest that our results may justify the future efforts in
studying generalized versions of the active advection models: inclusion of another new terms (like the KPZ interaction and its numerous anisotropic versions), complex coupling constants and transport parameters, 
non-local in-space and/or in-time random forces, {\it etc}. 
This work remains for the future and partly is already in progress. 

\section*{Acknowledgements}
The authors thank L.Ts.~Adzhemyan, N.M.~Gulitskiy, P.I.~Kakin, N.M.~Lebedev 
and M.~Yu.~Nalimov for discussions.
N.V.A. was supported by the Foundation for the Advancement of Theoretical Physics and Mathematics ``BASIS,''
grant 18-1-1-53-1.
M.M.K. was supported by the RFBR, project number
19-32-60065.

\end{document}